\documentclass[journal]{IEEEtran}
\usepackage{cite}

\usepackage{graphicx}

\begin{document}
\newcommand{\cm}[1]{\ensuremath{ {{\rm cm}^{#1}}}}
\newcommand{\ntp}[2]{\ensuremath{#1\times10^{#2} } }
\newcommand{\asb}[2]{\ensuremath{#1_{\rm #2} }}

\begin{titlepage}

\thispagestyle{empty}
\def\thefootnote{\fnsymbol{footnote}}       

\begin{center}
\mbox{ }

\end{center}
\begin{center}
\vskip 1.0cm
{\Huge\bf
Radiation Hardness Studies in a CCD with
}
\vspace{2mm}

{\Huge\bf
High-Speed Column Parallel Readout
}
\vskip 1cm
{\LARGE\bf 
Andr\'e Sopczak
}

\smallskip

\Large 
Lancaster University, UK; on behalf of the LCFI Collaboration

\vskip 2.5cm
\centerline{\Large \bf Abstract}
\end{center}

\vskip 4.cm
\hspace*{-0.5cm}
\begin{picture}(0.001,0.001)(0,0)
\put(,0){
\begin{minipage}{\textwidth}
\Large
\renewcommand{\baselinestretch} {1.2}
Charge Coupled Devices (CCDs) have been successfully used in several high energy physics 
experiments over the past two decades. Their high spatial resolution and thin sensitive 
layers make them an excellent tool for studying short-lived particles. The Linear Collider 
Flavour Identification (LCFI) collaboration is developing Column-Parallel CCDs (CPCCDs) 
for the vertex detector of the International Linear Collider (ILC). The CPCCDs can be 
read out many times faster than standard CCDs, significantly increasing their operating speed. 
The results of detailed simulations of the charge transfer inefficiency (CTI) of a prototype 
CPCCD are reported and studies of the influence of gate voltage on the CTI described. 
The effects of bulk radiation damage on the CTI of a CPCCD are studied by simulating the 
effects of two electron trap levels, 0.17 and 0.44~eV, at different concentrations 
and operating temperatures. The dependence of the CTI on different occupancy levels 
(percentage of hit pixels) and readout frequencies is also studied. The optimal operating 
temperature for the CPCCD, where the effects of the charge trapping are at a minimum, 
is found to be about 230~K for the range of readout speeds proposed for the ILC. 
The results of the full simulation have been compared with a simple analytic model.
\renewcommand{\baselinestretch} {1.}

\normalsize
\vspace{3.5cm}
\begin{center}
{\sl \large
Presented at the IEEE 2007 Nuclear Science Symposium, Honolulu, USA, \\
the Joint Meeting of the American Linear Collider Physics Group and ILC Global Design Effort, \\
and the 10th ICATPP Conference on Astroparticle, Particle, Space Physics, Detectors and \\
Medical Physics Applications, Como, Italy, to be published in the proceedings.
\vspace{-6cm}
}
\end{center}
\end{minipage}
}
\end{picture}
\vfill

\end{titlepage}

\newpage
\thispagestyle{empty}
\mbox{ }
\newpage
\setcounter{page}{0}

\title{Radiation Hardness Studies in a CCD with
       High-Speed Column Parallel Readout}

\author{
\authorblockN{Andr\'e Sopczak$^1$\thanks{$^1$Presented on behalf of the LCFI Collaboration; E-mail: andre.sopczak@cern.ch}\authorrefmark{1}, 
Salim Aoulmit\authorrefmark{5}, 
Khaled Bekhouche\authorrefmark{5}, 
Chris Bowdery\authorrefmark{1}, 
Craig Buttar\authorrefmark{6}, 
Chris Damerell\authorrefmark{3},\\
Gavin Davies\authorrefmark{1}, 
Dahmane Djjendaoui\authorrefmark{5}, 
Lakhdar Dehimi\authorrefmark{5}, 
Tim Greenshaw\authorrefmark{4}, 
Michal Koziel\authorrefmark{1},  
Dzmitry Maneuski\authorrefmark{6},\\
Konstantin Stefanov\authorrefmark{3}, 
Tuomo Tikkanen\authorrefmark{4},
Tim Woolliscroft\authorrefmark{4}, 
Steve Worm\authorrefmark{3}\\
\bigskip
}
\authorblockA{\authorrefmark{1}Lancaster University, UK\\}
\authorblockA{\authorrefmark{5}Biskra University, Algeria\\}
\authorblockA{\authorrefmark{6}Glasgow University, UK\\}
\authorblockA{\authorrefmark{3}STFC Rutherford Appleton Laboratory, UK\\}
\authorblockA{\authorrefmark{4}Liverpool University, UK\\}}

\maketitle

\begin{abstract}
Charge Coupled Devices (CCDs) have been successfully used in several high energy physics 
experiments over the past two decades. Their high spatial resolution and thin sensitive 
layers make them an excellent tool for studying short-lived particles. The Linear Collider 
Flavour Identification (LCFI) collaboration is developing Column-Parallel CCDs (CPCCDs) 
for the vertex detector of the International Linear Collider (ILC). The CPCCDs can be 
read out many times faster than standard CCDs, significantly increasing their operating speed. 
The results of detailed simulations of the charge transfer inefficiency (CTI) of a prototype 
CPCCD are reported and studies of the influence of gate voltage on the CTI described. 
The effects of bulk radiation damage on the CTI of a CPCCD are studied by simulating the 
effects of two electron trap levels, 0.17 and 0.44~eV, at different concentrations 
and operating temperatures. The dependence of the CTI on different occupancy levels 
(percentage of hit pixels) and readout frequencies is also studied. The optimal operating 
temperature for the CPCCD, where the effects of the charge trapping are at a minimum, 
is found to be about 230~K for the range of readout speeds proposed for the ILC. 
The results of the full simulation have been compared with a simple analytic model.
\end{abstract}

\section{Introduction}
Particle physicists worldwide are working on the design of a high
energy collider of electrons and positrons (the International Linear
Collider or ILC) which could be operational sometime around 2019. Any
experiment exploiting the ILC will require a high performance vertex
detector to detect and measure short-lived particles, yet be
tolerant to radiation damage for its anticipated lifetime. One candidate
is a set of concentric cylinders of
Charge-Coupled Devices (CCDs), read out at a frequency of around 50\,MHz.

It is known that CCDs suffer from both surface and bulk radiation damage. However,
when considering charge transfer losses in buried channel devices
only bulk traps are important. These defects create energy levels (traps)
between the conduction and valence band, and electrons are
captured by them. These electrons are also emitted back
to the conduction band after a certain time.

\begin{figure}[htp]
\includegraphics[width=\columnwidth]{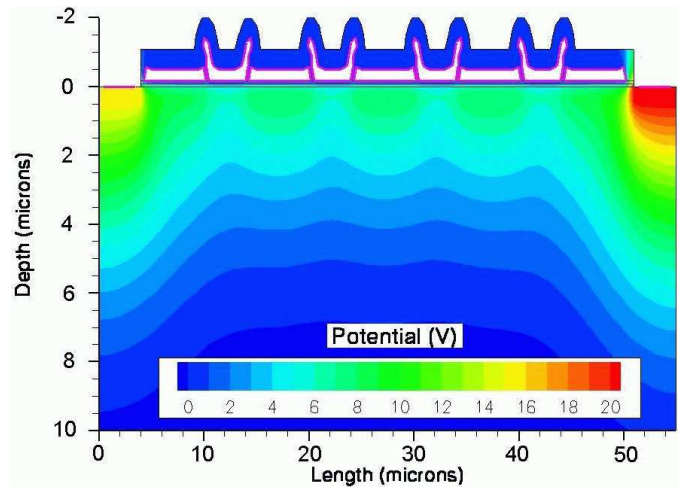}
\begin{center}
\begin{tabular}{c | c| c}
$E (eV)$ 
 & $C$ (\cm{-3}) &$\sigma_{\rm n}$ (\cm{2})\\ \hline
0.17 & \ntp{1}{12} & \ntp{1}{-14}\\
0.44 & \ntp{1}{12} & \ntp{1}{-15}
\end{tabular}
\end{center}
\caption{\label{fig:trap}
Upper: detector structure and potential at gates (nodes) after initialization. 
The pixel is located around between 20 and 40 $\rm \mu m$ length.
Lower: energy levels $E$, trap concentrations $C$, and electron-capture cross-section $\sigma_{\rm n}$ 
used in simulation. }
\end{figure}

It is usual to define the Charge Transfer Inefficiency
(CTI) as the fractional loss of charge after transfer across one pixel.
An initial charge $Q_0$ after being transported across
$m$ pixels is reduced to
$Q_m=Q_0(1-{\rm CTI})^m$.
For CCD devices containing many pixels, CTI values around 10$^{-5}$ are important.
Previous results for a CCD with sequential readout have recently been reported~\cite{ieee}.
The expected background rate at the future ILC near the interaction point leads to
radiation damage in the CCD detector.
We simulated the charge transfer in a CPCCD (Column-Parallel CCD) using the trap 
concentrations listed in Fig.~\ref{fig:trap}. 
They correspond to about 3 years of operation for the 0.17 eV traps and 
several more years for the 0.44 eV traps.

\section{Simulations}

The UK Linear Collider Flavour Identification (LCFI)
collaboration~\cite{LCFI} has been studying a CCD with high-speed (50 MHz) 
column-parallel readout produced by e2V Technologies. 
It is a two-phase buried-channel CCD with 20\,$\mu$m square pixels.

Simulations of a simplified model of this device have been performed with
the ISE-TCAD package (version 7.5), particularly the DESSIS program
(Device Simulation for Smart Integrated Systems).
The simulation is essentially two dimensional and assumes a 1\,$\mu$m 
device thickness (width) for calculating densities.
The overall length and depth are 55\,$\mu$m and 20\,$\mu$m respectively
(Fig.~\ref{fig:trap}).

\begin{figure}[htp]
\includegraphics[width=\columnwidth,clip]{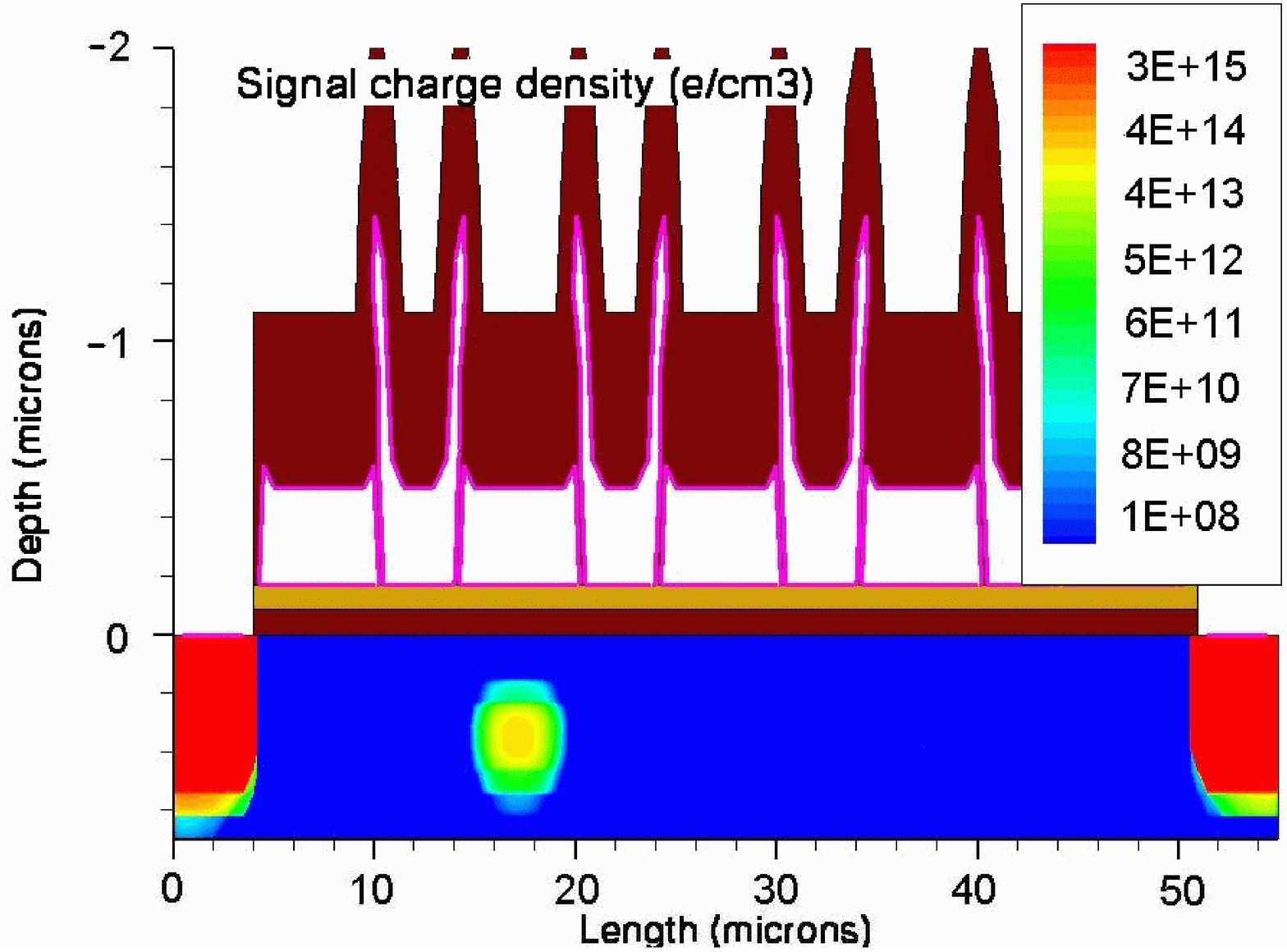}
\includegraphics[width=\columnwidth,clip]{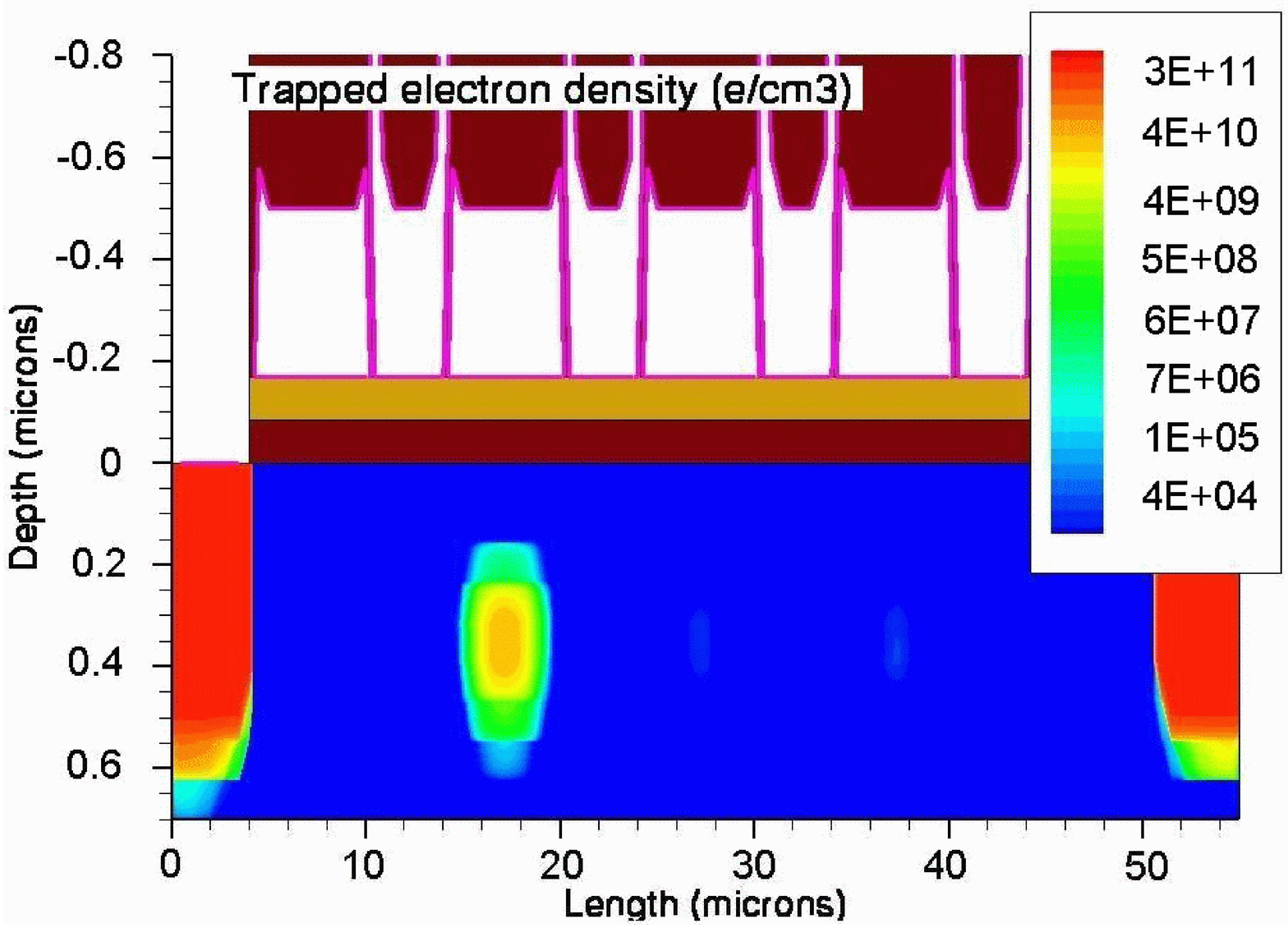}
\caption{\label{fig:transport} 
Upper: signal charge density.
Lower: trapped charge density.
}
\end{figure}

Parameters of interest are the readout frequency, up to 50\,MHz, and
the operating temperature between 130\,K and 300\,K although
simulations have been performed up to 440\,K. The charge in transfer and
the trapped charge are shown in Fig.~\ref{fig:transport}.

Charge Transfer Inefficiency is a measure of the fractional loss of
charge from a signal packet as it is transferred over a pixel, or
two gates. After DESSIS has simulated the transfer process, a 2D
integration of the trapped charge density distribution is performed
independently to give the total charge under each gate.

This CTI study, at nominal clock voltage, focuses only on
the bulk traps with energies 0.17\,eV and 0.44\,eV below the bottom
of the conduction band. These will be referred to simply as the
0.17\,eV and 0.44\,eV traps. 
The 0.17~eV trap is an oxygen-vacancy defect,
referred to as an A-centre defect. The 0.44~eV trap is a phosphorus-vacancy 
defect, an E-centre defect, as a result of the silicon being
doped with phosphorus and a vacancy manifesting from the displacement of a
silicon atom bonded with a phosphorus atom~\cite{Stefanov}.

\section{Simulation Results}

\begin{figure}[hbp]
\vspace*{5mm}
\includegraphics[width=\columnwidth,clip]{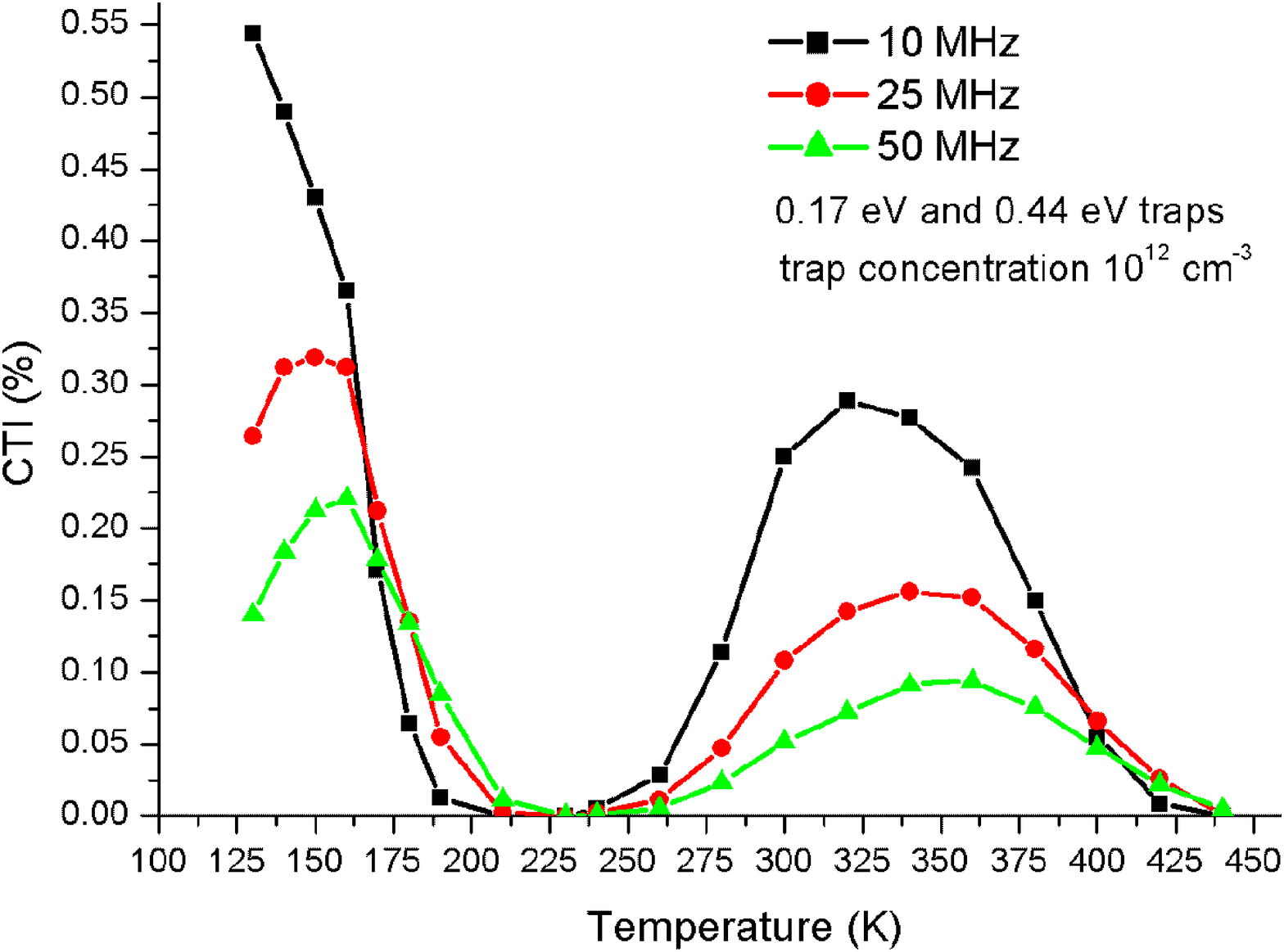}

\vspace*{10mm}
\includegraphics[width=\columnwidth,clip]{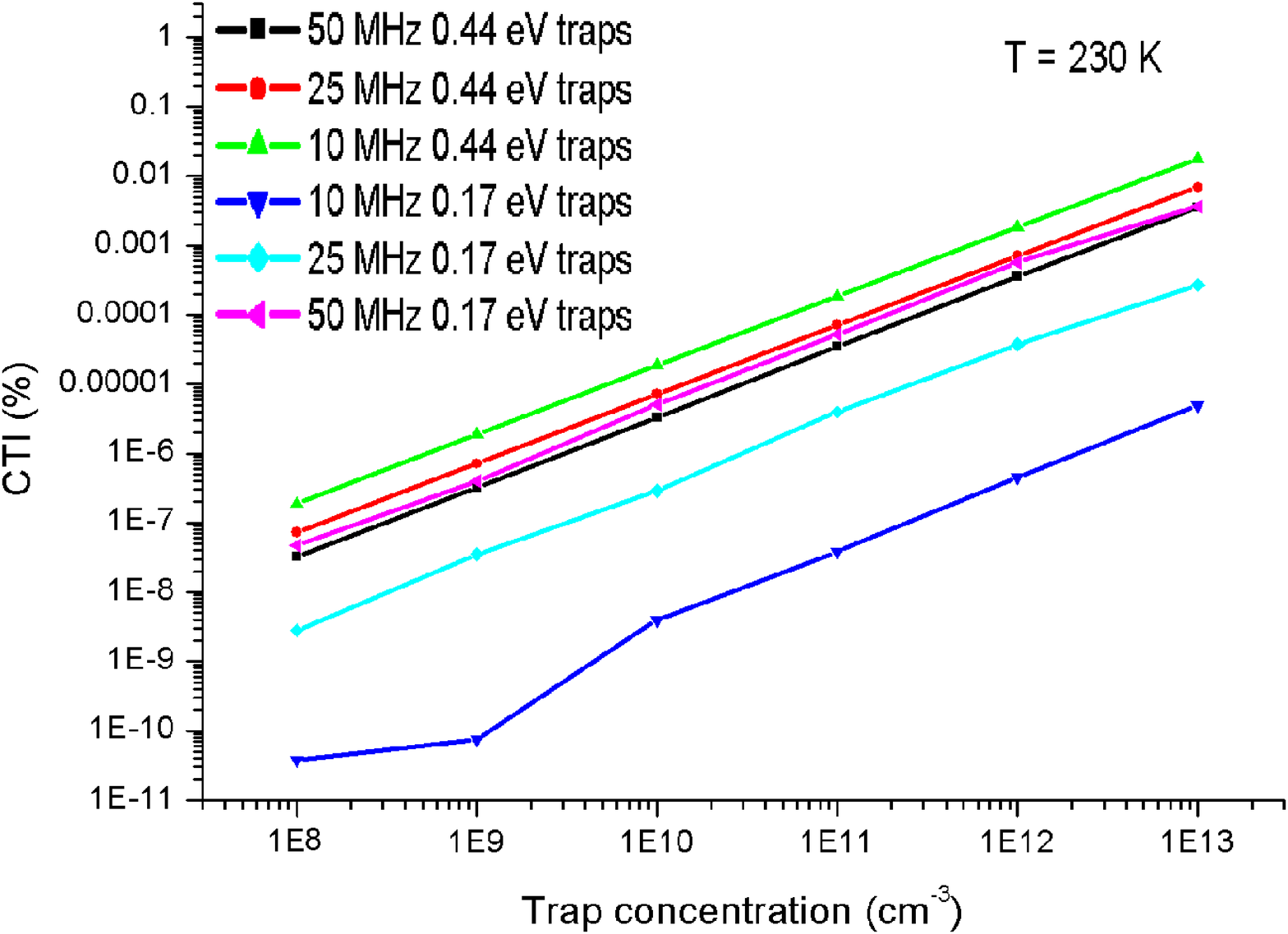}
\caption{\label{fig:allfreq_0_17eV}
Upper: CTI values for 0.17 and 0.44~eV traps at clocking frequencies 10, 25 and 50~MHz.
Lower: CTI values for a large range of trap concentrations.
}
\end{figure}

The CTI dependence on temperature and readout frequency was explored and
Figure~\ref{fig:allfreq_0_17eV} shows the CTI for simulations with
partially filled 0.17\,eV and 0.44\,eV traps at different frequencies for
temperatures between 130\,K and 440\,K, with a nominal clock voltage
of 3\,V. The CTI value depends linearly on the trap concentration for
a large concentration variation as also shown in Fig.~\ref{fig:allfreq_0_17eV}.

\begin{figure}[htp]
\includegraphics[height=6cm,width=\columnwidth,clip]{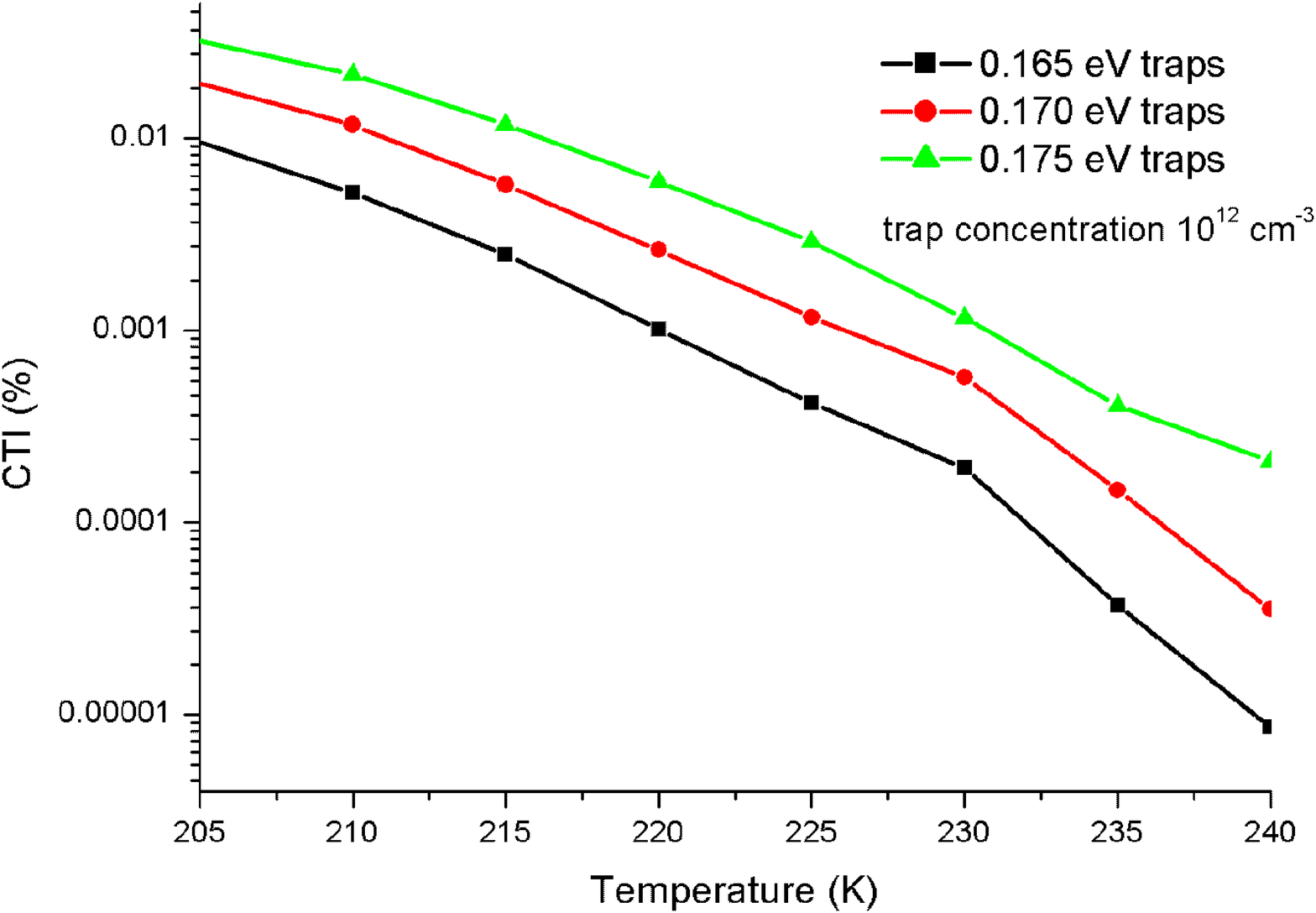} 

\vspace*{1mm}
\includegraphics[height=6cm,width=\columnwidth,clip]{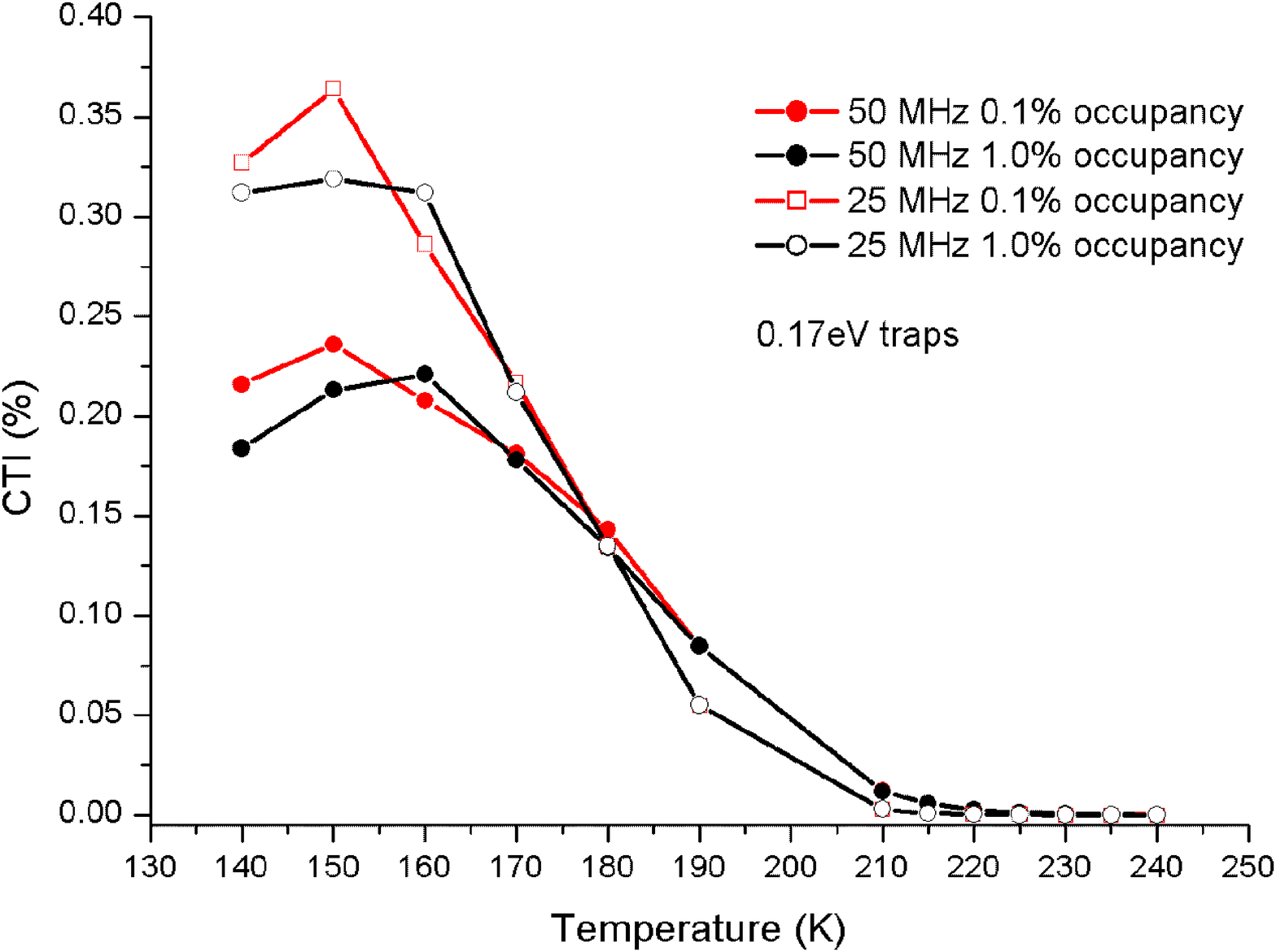}

\vspace*{-2mm}
\caption{\label{fig:energy}
Upper: CTI values for 0.165, 0.170 and 0.175~eV traps. 
Lower: CTI values for 0.1\% and 1\% hit (pixel) occupancy.
}
\vspace*{-1.5mm}
\end{figure}

\begin{figure}[htp]
\includegraphics[height=4.5cm,width=\columnwidth,clip]{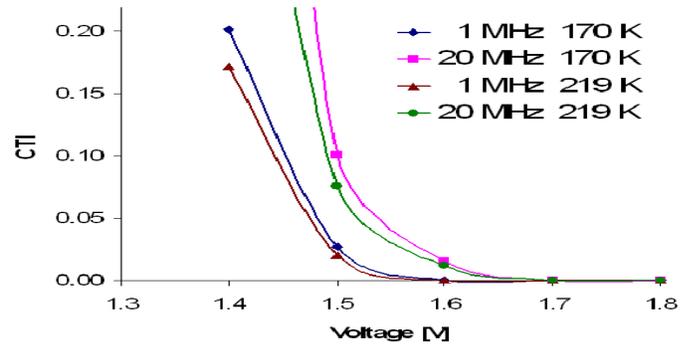} 

\vspace*{-2mm}
\caption{\label{fig:ISE_v_IM} 
Clock voltage induced CTI.}
\end{figure}

The CTI values for a variation of the 0.17~eV trap level by 0.005~eV and
for 0.1\% and 1\% hit (pixel) occupancy are shown in Fig.~\ref{fig:energy}.
Figure~\ref{fig:ISE_v_IM} shows the results of a study of a clock 
voltage induced CTI in order to find the optimum clock voltage
with a low power consumption and keeping the CTI at an acceptable level.

\section{Comparisons with an Analytic Model}
\begin{figure}[htp]
\includegraphics[height=5cm,width=\columnwidth,clip]{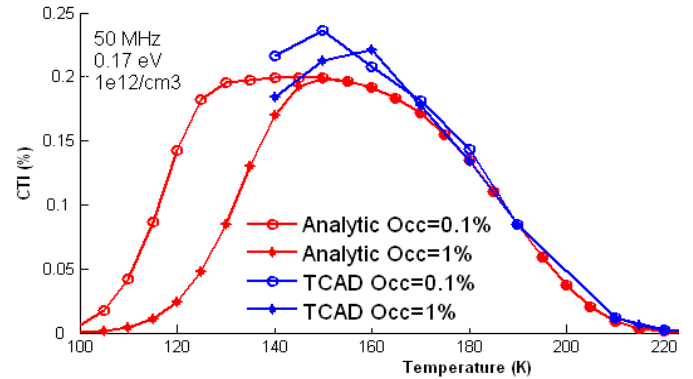}

\vspace*{-2mm}
\caption{\label{fig:ISE_v_IM_comp} 
CTI values for 0.17~eV traps at clocking frequency 50\,MHz. 
Comparison of the analytic model with the full TCAD simulation 
for two hit (pixel) occupancies of 0.1\% and 1\%.}
\end{figure}

The motivation for introducing an analytic model is to understand the underlying 
physics through making comparisons with the TCAD simulations. This might then allow 
predictions of CTI for other CCD geometries without requiring a full simulation.
A simple analytic model~\cite{ieee} has been adapted to the CPCCD characteristics.
The analytic model gives very similar results as shown in Figs.~\ref{fig:allfreq_0_17eV}
and~\ref{fig:energy}.
Figure~\ref{fig:ISE_v_IM_comp} compares the full TCAD simulation for 0.17\,eV traps and 
clocking frequency of 50\,MHz to the analytic model. It emphasises the good agreement 
between the model and full simulations\footnote{For temperatures below 140K the full simulation does not converge.}. 
The variation of hit occupancy is studied and as expected at lower temperatures larger hit occupancy leads to 
lower CTI values as traps are filled to a higher level.

However, there are limitations with the analytic model.
They could relate to a breakdown of the assumptions at high temperatures,
to ignoring the precise form of the clock voltage waveform, or to
ignoring the pixel edge effects.

\vspace*{-2mm}
\section{Conclusions and Outlook}

The Charge Transfer Inefficiency (CTI) of a CCD with high-speed column parallel readout 
has been studied with a full simulation (ISE-TCAD DESSIS) and compared with an analytic model.
The 0.17\,eV and 0.44\,eV trap levels have been implemented in the full simulation and variations 
of the CTI with respect to temperature and frequency have been analysed.
At low temperatures ($<230$\,K) the 0.17\,eV traps
dominate the CTI, whereas the 0.44\,eV traps dominate at higher temperatures.
Good agreement between simulations and an analytic model adapted to the CPCCD has been found. 
The optimum operating temperature for the CPCCD prototype in a high radiation
environment is found to be about 230\,K for clock frequencies in the range 10 to 50\,MHz.
Our prototype CPCCD has recently operated at 45\,MHz and a test-stand for CTI measurements is
in preparation. The development of a high-speed CCD vertex detector is on track as a vital part
of a future ILC detector.

\vspace*{-1mm}
\section*{Acknowledgments}
\vspace*{-1mm}
This work is supported by the Science and Technology Facilities Council (STFC)
and Lancaster University. The Lancaster authors wish to thank Alex Chilingarov, 
for helpful discussions, and the particle physics group at Liverpool University, for the 
use of its computers.

\end{document}